\documentclass{article}
\usepackage{spconf,amsmath,graphicx}

\title{DEEP-LEARNING-BASED MAGNETIC RESONANCE SIMULTANEOUS MULTISLICE IMAGING USING HOLOGRAPHIC IMAGE DECODING}
\twoauthors
  {Satoshi ITO\sthanks{This work was partially supported by the Japan Society for the Promotion of Science KAKENHI under grant 23K03864, the Suzuken Memorial Foundation, and the Hagiwara Foundation of Japan.}, Yuki SATO, Naoya ENDO}
	{Graduate Program in Information, Electrical\\
	 and Electronic Systems Engineering\\
	Utsunomiya University}
{Shohei OUCHI}
	{Department of Innovative Electrical\\
    and Electronic Engineering\\
	National Institute of Technology, Oyama}
\begin{document}
%
\maketitle
\begin{abstract}
Simultaneous multislice (SMS) imaging is a one of the acceleration technique of magnetic resonance imaging. 
SMS requires accurate sensitivity distributions in the slice plane for each receiving coil. This requirement is difficult to satisfy in practice, limiting the applications of this imaging technique.
Here, images are reconstructed by applying deep learning and amplitude modulation to each slice image.
Simulation experiments show that image reconstruction can be achieved for both real- and complex-valued images. Image quality tends to decrease with increasing number of simultaneously acquired images. It is also shown that a larger difference in the phase modulation coefficients between slices tends to increase the quality of the reconstructed images.
Simulation experiments and initial MR imaging experiments show promising results for this method.
\end{abstract}
%
\begin{keywords}
simultaneous multislice imaging, holography, reconstruction, convolutional neural network (CNN)
\end{keywords}
\section{Introduction}
\label{sec:intro}

Among acceleration methods for magnetic resonance imaging (MRI), such as compressed sensing, parallel imaging, and simultaneous multislice (SMS) imaging, SMS imaging has the advantage of speeding up imaging in proportion to the number of slices excited simultaneously without signal undersampling.
The simultaneous excitation and imaging of several slices was first realized in the 1980s~\cite{1-1,1-2}.
Phase-offset multiplanar volume imaging (POMP) develop the SMS imaging~\cite{1-3}, 
POMP extends the field of view and increases the number of signal acquisitions in the phase-encoding direction, but does not speed up imaging.

The invention of parallel imaging~\cite{1-4} has greatly advanced SMS techniques and applications.
SMS imaging typically uses the sensitivity distribution of the receiver radio-frequency (RF) coils to separate overlapping slice images. Therefore, accurate sensitivity distributions at the slice position are required for each receiver coil.
If the slice positions are close to each other, it is difficult to separate the slice images due to the small difference in the sensitivity distributions of the receiver RF coils at the slice positions.
SMS imaging thus has relatively few application at that time.
CAIPIRINHA (Controlled Aliasing in Parallel Imaging Results in Higher Acceleration)~\cite{1-5} has greatly accelerated the use of SMS imaging. The CAIPIRINHA method, which does not use standard Cartesian sampling, reduces image overlap, resulting in higher image quality.

We previously applied the principle of holography to MR imaging method to reconstruct an arbitrarily focused image at the $z$-coordinate from two-dimensional signals~\cite{1-6,1-7}. The challenge of this imaging method is eliminating out-of-focus images.
 In the present study, we apply deep learning to separate superimposed images.
In addition, for practical applications, we consider a format that can be applied to the Fourier transform imaging method used with MRI in clinical use.
Even though the SMS concept is incorporated into the proposed method, image separation is conducted without using the sensitivity distribution of the receiver RF coils.

The major difference between the proposed method and conventional methods is that the proposed method uses deep learning for image separation and does not use the sensitivity distributions of the receiver coils. 
Although the proposed method requires a long learning time, it greatly reduces hardware requirements and does not require solving simultaneous equations. Higher image quality is thus expected.
The reconstruction performance of the proposed method is investigated using simulation experiments as a preliminary study and initial reconstruction experiments using low-field MRI.
%
\section{SIMULTANEOUS multislice IMAGING USING MR HOLOGRAPHIC IMAGING}
\label{sec:format}

We previously proposed an MR imaging method that incorporates the principle of holography~\cite{1-6,1-7}. This section briefly describes the signal equations for Fresnel-holography-based imaging and the proposed Fourier-holography-based SMS method.
\subsection{Fresnel-holography-based MR imaging}
The imaging technique based on Fresnel holography encodes the spatial information of the subject using a diffraction equation for light or sound waves. A quadratic field gradient, whose field intensity changes in a quadratic form on the $x-y$ plane and whose coefficient varies in the $z$ direction by $\alpha$, is scanned in two dimensions in the $x-y$ plane. The MR signal equation $v(x', y')$ is given as~\cite{1-7}

\vspace{-1.5mm}
\footnotesize
\begin{eqnarray}
&&\hspace{-8mm} v(t, y')  \!=\! e^{-j \gamma b_0 t } \! \int \hspace{-1.5mm}  \int \hspace{-2.8mm}  \int\limits_{\,\,\,-\infty}^{\,\,\,\,\,\,\infty} \! \rho(x,y,z)   e^{-j\gamma b \tau \left(1+\alpha z\right) \left[ (x'-x)^{2} + y^2 \right] } \nonumber \\
%
%
&&\hspace{-3mm} \cdot e^{-j \gamma \left(1\!+\!\alpha z\right) G_y t\!+\!\gamma \int_{-t_{0}}^{t} \!\! \left(1\!+\!\alpha z\right) b_{0y} t dt } dxdydz \nonumber \\ 
&&\hspace{-6mm} = e^{(-j \gamma b_0 t+\frac{\gamma b_{0y}}{2} {t_0}^2 ) }\! \int \hspace{-1.5mm} \int \hspace{-2.8mm} \int\limits_{\,\,\,-\infty}^{\,\,\,\,\,\,\infty} \!\! \rho (x,\!y,\!z) e^{-j\gamma b \tau \left(1\!+\!\alpha z\right) \left[ (x'\!-\!x)^2\!+\! \left(y\!+\!\frac{G_y t}{2 b \tau} \right)^2 \right] }  \nonumber \\
&& \hspace{-2mm} \cdot e^{j \gamma t^2 \left(1+\alpha z\right) \left(\frac{{G_y}^2}{4 b \tau}-\frac{b_{0y}}{2}\right) }dxdydz ,
\label{eqn : N3d_fresnel}
\end{eqnarray}
\normalsize
%
\begin{figure}[t]
\centering
 \centerline{\includegraphics[width=7.0cm]{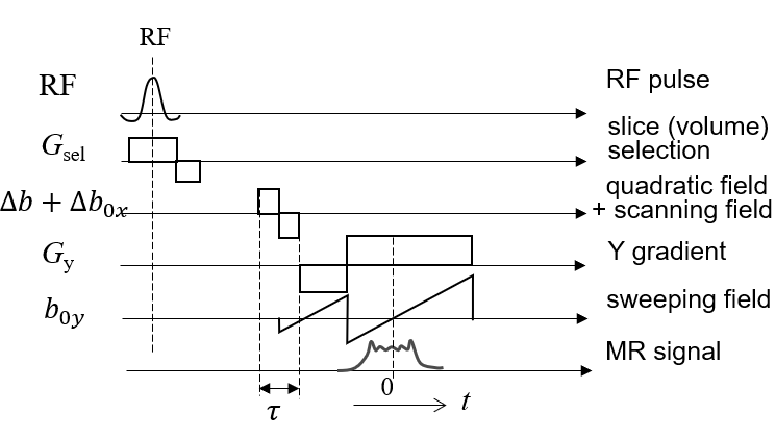}}
%
\caption{\small Pulse sequence for Fresnel-holography-based MR imging.}
\label{fig:FRpulse}
\end{figure}

\noindent
where $\rho(x,y)$ is the spin density distribution for the subject, $\gamma$ is the gyromagnetic ratio, $\tau$ is its impressed time, $b$ is a coefficient of the quadratic field gradient at $z=0$, and coordinates $(x',y')$ are the center of the quadratic field gradient. Figure 1 shows the pulse sequence used in this method. The coordinates can be set via an external field (i.e., scanning field $b_{0x}$ and sweeping field $b_{0y}$ for the $x$ and $y$ directions, respectively). Setting the parameters to satisfy {\small ${G_y}^2/4 b \tau=b_{0y}/2$ } and replacing variables as {\small $y'=-G_y t/2 b \tau$ } and {\small $P=\exp{(-j \gamma b_0 t+\frac{\gamma b_{0y}}{2} {t_0}^2 ) }$ } yields

\footnotesize
\begin{equation}
v(x',y')\! =\! \! P \! \int \hspace{-1.5mm} \int \hspace{-1.5mm} \int^{\infty}_{-\infty} \hspace{-1.0mm} \{ \rho(x,y,z) e^{-j \gamma b \tau (1+\alpha z) \{(x'-x)^2+(y'-y)^2 \}} dxdydz .
\end{equation}
\normalsize

\noindent
Images focused at the $z_i$-plane can be reconstructed using the inverse filtering technique:

\vspace{-1.5mm}
\footnotesize
\begin{equation}
\rho(x,y,z_i)\! =\! \frac{1}{P} \sqrt{ \frac{\gamma b \tau (1\!+\!z')}{\pi} } {\cal F}^{-1} \! \left[e^{j \frac{\pi}{4}} e^{-j \frac{ {k_x}^2+{k_y}^2}{4 \gamma b \tau (1+z')} } {\cal F} \left[ v(x',y') \right] \right] .
\end{equation}
\normalsize

\vspace{-6mm}
\subsection{Fourier-holography-based SMS imaging}
Considering practical applications, SMS imaging is discussed here. The case where three slice images $(z_1, z_2, z_3)$ are acquired simultaneously is considered. Figure 2 shows the pulse sequence used in this method. Although it is possible to excite all slices with a single RF pulse in principle, here, we apply a simple and low-RF-peak-power method. After RF pulse 1 is applied to excite only slice 1, phase modulation is applied to slice 1 by applying gradient $G_{y1}$ for time $t_1$. Next, RF pulse 2 is applied to excite slice 2 and gradient $G_{y2}$ is applied to this slice for time $t_2$ to apply phase modulation to slices 2 and 1. Slice 3 is phase-modulated in the same way. After multiple rounds of slice excitation and phase modulation, signal acquisition is performed using a Fourier transform imaging sequence. The MR signal in this sequence is expressed as

\vspace{-1.5mm}
\footnotesize
\begin{eqnarray}
&&\hspace{-9mm} v(k_x,k_y)\!= \!\! \int \hspace{-1.5mm} \int^{\infty}_{-\infty} \hspace{-1.5mm} \{ \rho(x,y,z_1) e^{-j (k_{1}\!+\!k_{2}\!+\!k_{3}) y}\!+\!\rho(x,y,z_2) e^{-j (k_{2}+k_{3}) y}  \nonumber \\
&&\hspace{4mm} +\rho(x,y,z_3) e^{-j k_{3} y} \} e^{-j \left(k_x x+k_y y \right)} dxdy
\end{eqnarray}
\begin{figure}[t]
\centering
 \centerline{\includegraphics[width=8.5cm]{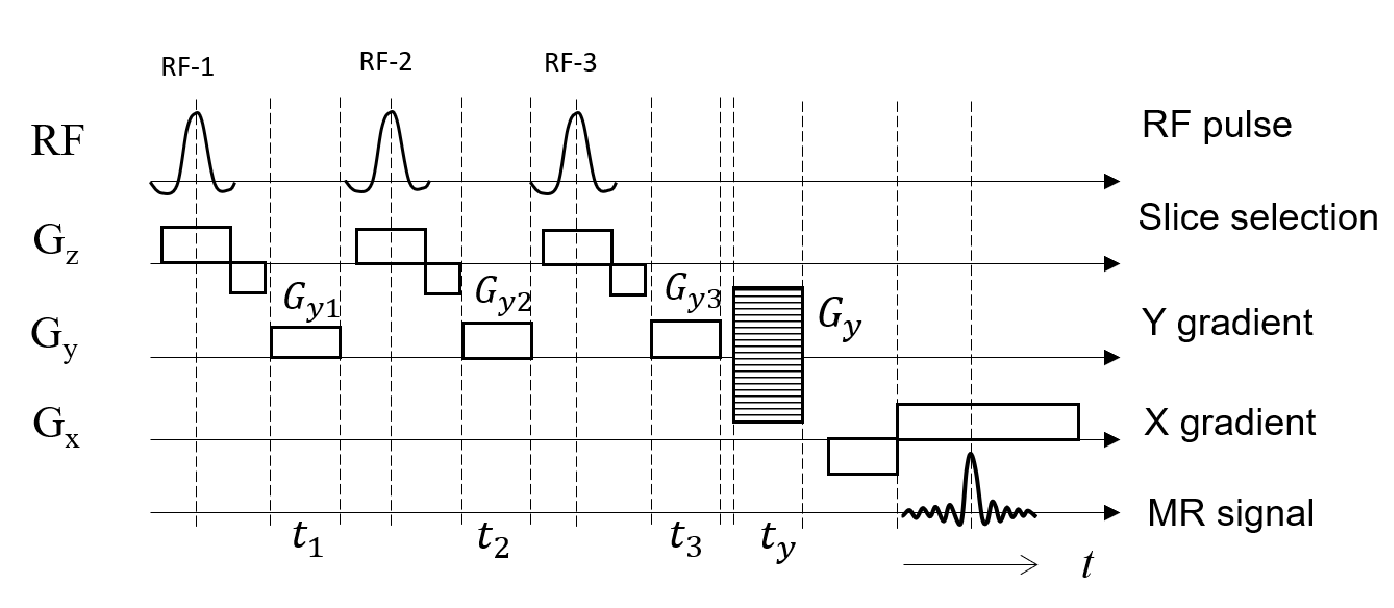}}
%
\renewcommand{\baselinestretch}{0.25}
\caption{\small Pulse sequence for Fourier-holography-based SMS imaging. Slice 1 is excited by RF pulse 1 and amplitude modulation is applied by gradient $G_{y1}$. Next, slice 2 is excited by RF pulse 2 and amplitude modulation is applied to slice 2 by gradient $G_{y2}$. At the same time, phase modulation is added to the already excited slice 1. The same process is repeated for other slices. Different amplitude modulation can be applied to each slice.}
\label{fig:pulse}
\end{figure}

\normalsize
\noindent
where $k_{1}= \gamma G_{y1} t_1$, $k_{2}= \gamma G_{y2} t_3$, and $k_{3}= \gamma G_{y3} t_3$.
The image focused on the $z_1$ plane $\rho_{t}(x,y,z_1)$ is obtained by compensating for the phase modulation term applied to the $z_1$ plane in the inverse-Fourier-transformed image:

\vspace{-1.5mm}
\footnotesize
\begin{eqnarray}
&&\hspace{-13mm} \rho_{t}(x,y,z_1) = e^{j (k_{1}+k_{2}+k_{3}) y}  {\cal F}^{-1} \left[v(k_x,k_y) \right] \nonumber \\
&&\hspace{-6mm} =\! \rho(x,y,z_1)\! +\! \rho(x,y,z_2) e^{j k_{1} y} \!+\!\rho(x,y,z_3) e^{j (k_{1}+k_{2}) y}
\end{eqnarray}
\normalsize

\noindent
In Eq.~(5), only focal-plane image $\rho(x,y,z_1)$ has a form without an exponential term. Temporary reconstructed image $\rho_{t}(x,y,z_1)$ is disturbed by out-of-focus  images.
\section{IMAGE SEPARATION USING U-NET}
\label{sec:sepimages}
Three images focused at the $z_1$, $z_2$, and $z_3$ planes are obtained by changing the exponential term in Eq.~(5). The expression for the obtained images is a simultaneous equation of rank 1, so it is not possible to algebraically solve the focal-plane images excluding the out-of-focus images.
Therefore, we extract the focal-plane images using a convolutional neural network (CNN). Figure 3 shows the scheme used for image separation with out-of-focus images.
Temporary reconstructed images are used as the input images and the focal-plane images are used as the teacher images to train the CNN. In our study, U-Net~\cite{3-1} was used as the network model.
Since the temporary reconstructed images are complex-valued, we performed CNN training for each slice image as an initial study. Therefore, the CNN used for reconstruction was different for each slice image.
\begin{figure}[t]
\centering
 \centerline{\includegraphics[width=8.0cm]{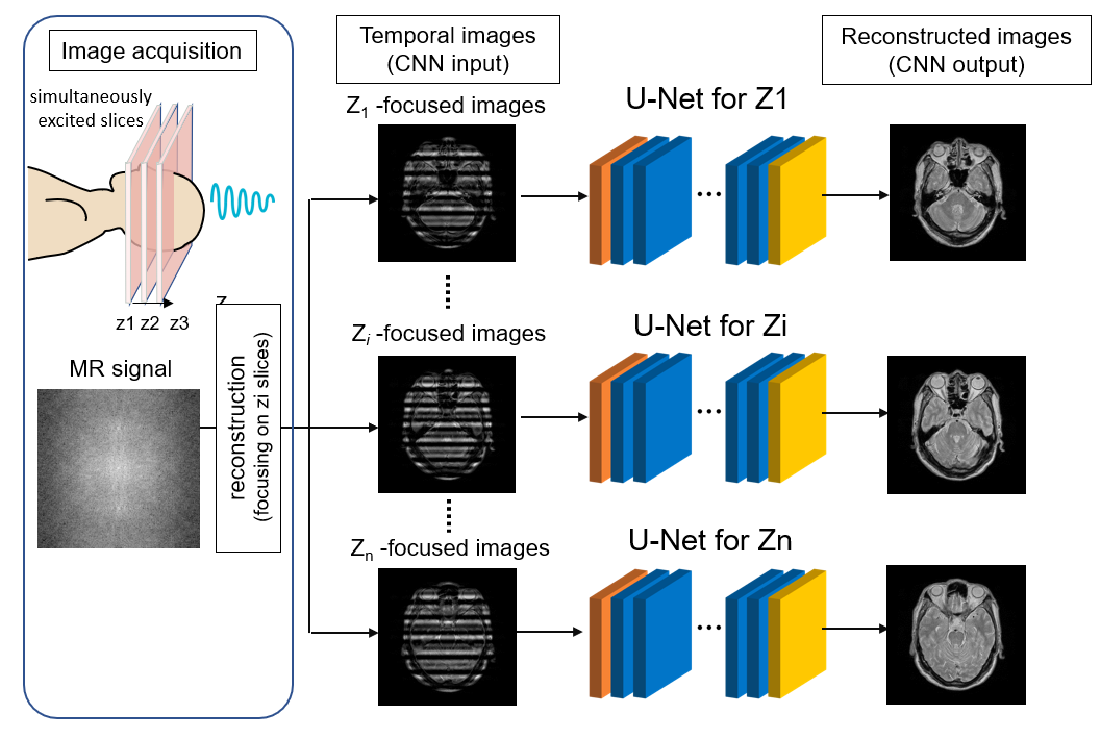}}
%
\vspace{-4mm}
\caption{\small Scheme used for  deep-learning-based SMS imaging. \newline Focused images at each slice position are used as the input images of the CNN and true slice images are used as teacher images.}
\label{fig:pulse}
\end{figure}

\section{EXPERIMENTS}
\label{sec:simulation}
\subsection{Dataset and experimental environment}
For training, 1400 T2-weighted and 1400 PD-weighted images from the IXI dataset were used (250 images for validation, 250 images for testing, $256\times256$ pixels, pixel size: $0.89\times0.89$ mm$^3$)~\cite{4-1}. Slices were spaced 3.75 mm apart.
For reconstruction experiments with complex-valued images, 400 T2-weighted images from the fastMRI dataset were used (100 images for validation, 100 images for testing, $320 \times 320$ pixels, pixel size: 0.5 mm $\times$ 0.5 mm)~\cite{4-2}. Slices were spaced 3 mm apart.
Since fastMRI is a parallel imaging dataset, the reconstructed images were used for training.
The $320\times320$-pixel fastMRI images were resized to $256\times256$-pixel images.
\begin{figure}[t]
\centering
 \centerline{\includegraphics[width=8.0cm]{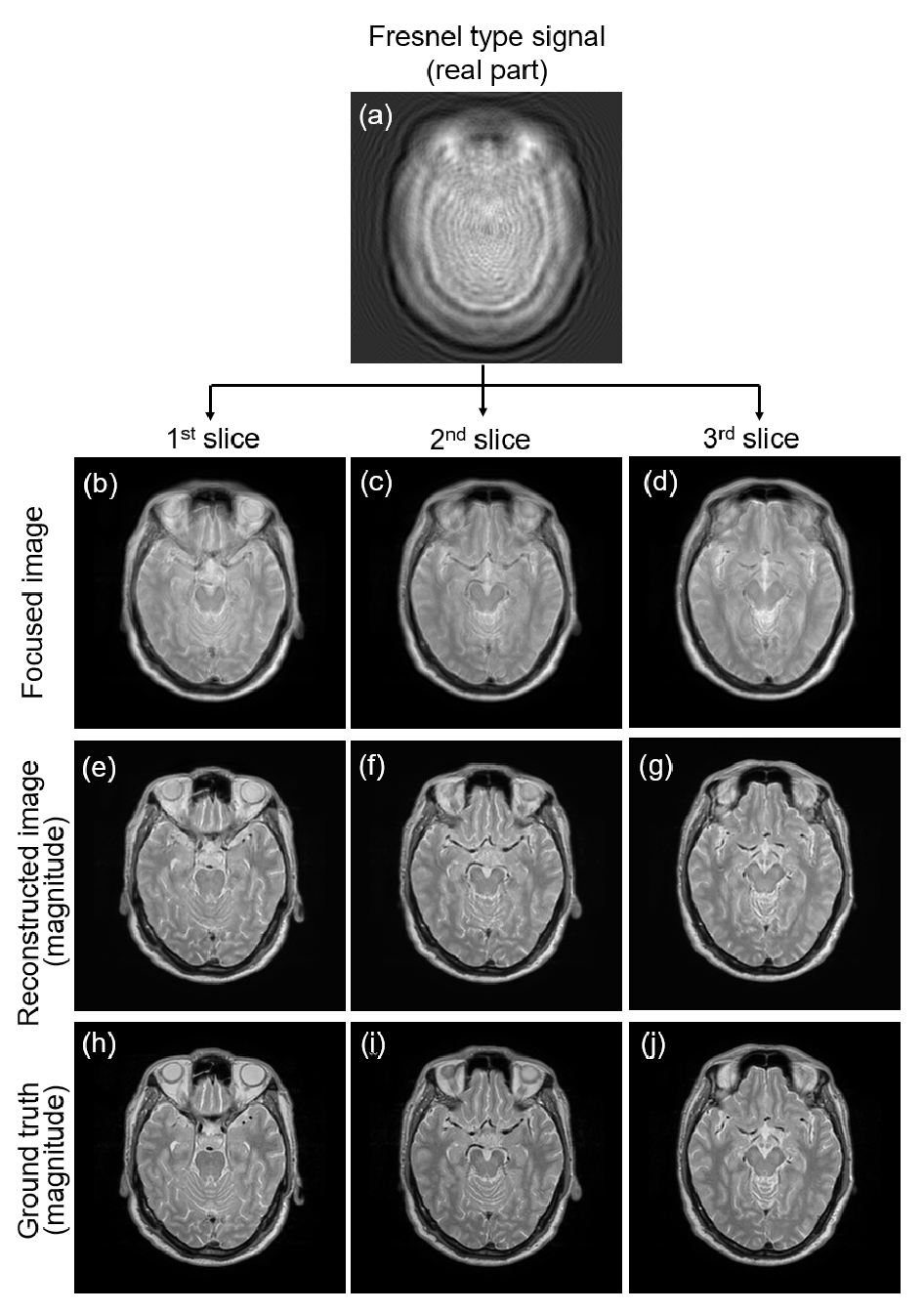}}
%
\vspace{-4mm}
\renewcommand{\baselinestretch}{0.3}
\caption{\small Results of Fresnel-hologram-based SMS imaging.
(a) MR signal obtained in Fresnel-type three-slice SMS imaging, (b)--(d) focused image at each slice position (CNN input), (e)--(g) images separated by CNN, and (h)--(j) true images corresponding to (e)--(g), respectively.}
\label{fig:pulse}
\end{figure}

We used MATLAB R2019b to implement the proposed method. The CNN architectures of the proposed method were constructed using the MatConvNet package~\cite{4-3}. PyTorch~\cite{4-4} was used for the complex-valued CNN. All CNN learning and reconstruction experiments were performed using a computer
with an Intel Core i7-9700K (3.60 GHz) CPU and an NVIDIA GeForce RTX 2080 Ti GPU. The proposed method was trained for 200 epochs using Adam as
the optimization algorithm.
\subsection{MR imaging conditions}
The imaging experiments were performed with our Fresnel-holography-based coil system designed for low-field MRI\cite{1-6}.
The MRI system generates a static magnetic field of 0.0183 T (resonant frequency is 779 kHz).
Single receiver RF coil was used for acquiring gradient echo signals.
In this experiment, volume excitation and imaging were performed, but was approximated as five-slice SMS (slice thickness 4 mm with no slice spacing) and reconstructed using a CNN pre-trained on 5 slices.
A small water pin phantom was used for imaging.
Imaging conditions were TR=300 ms, $\gamma b \tau=1.49  {\rm rad/cm^2 (\alpha = 0.033 /cm)}$. Scan width of quadratic field $\Delta x'=\Delta y'$=0.2 cm, spatial resolution $\Delta x=\Delta y=$0.2 cm.
Matrix size of signal is 64 $\times$ 64.
\subsection{Results of Fresnel-holography-based SMS imaging}
We now consider the normalization of the imaging parameter $\gamma b \tau$ in Eq.~(2). For simplicity, let us consider the $x$ direction. The number of data is $N$ and the pixel width is $\Delta x$. The value that satisfies the sampling theorem in the entire image space in the $x$ direction is given by $\overline{\gamma b \tau}=\pi/(N \Delta x^2)$. The imaging parameter $\gamma b \tau$ is expressed as $\gamma b \tau=h \overline{\gamma b \tau}$ using the coefficient $h$.
Simulation experiments of Fresnel-holography-based SMS images were conducted using three sliced images. The imaging parameters used for the three slice images were $h=0.8, 0.9, and 1.0$.
Figure 4 shows the results of the reconstruction experiments. Figure 4(a) shows the real part of the MR signal obtained using the slice images shown in Figs. 4(h)--(j). Figures 4(b)--(d) show the temporary reconstructed images focused at each slice position and used for the input image of each CNN.
Images separated by the CNN are shown in Figs. 4(e)--(g). Out-of-focus images were removed and focal-plane images were successfully reconstructed. The average peak signal-to-noise ratio (PSNR) and structural similarity index (SSIM) of these three images are 29.2 dB and 0.921, respectively.
\begin{figure}[t]
\centering
 \centerline{\includegraphics[width=9.0cm]{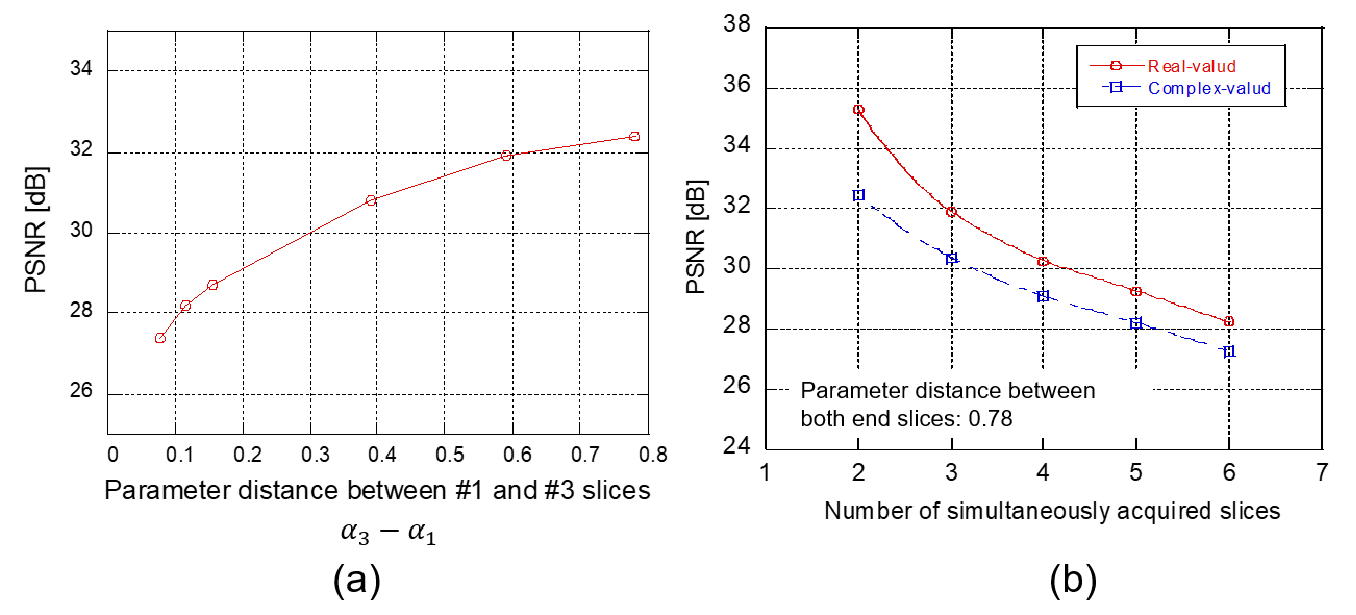}}
%
\vspace{-2mm}
\renewcommand{\baselinestretch}{0.3}
\caption{\small Relationship between PSNR and imaging parameters.  (a) relation between PSNR and difference in phase modulation coefficients between two end slices for three-slice SMS imaging. (b) Relationship between number of simultaneously imaged slices and average PSNR.}
\label{fig:slicenum}
%
%
\vspace{3mm}
\centering
 \centerline{\includegraphics[width=8.4cm]{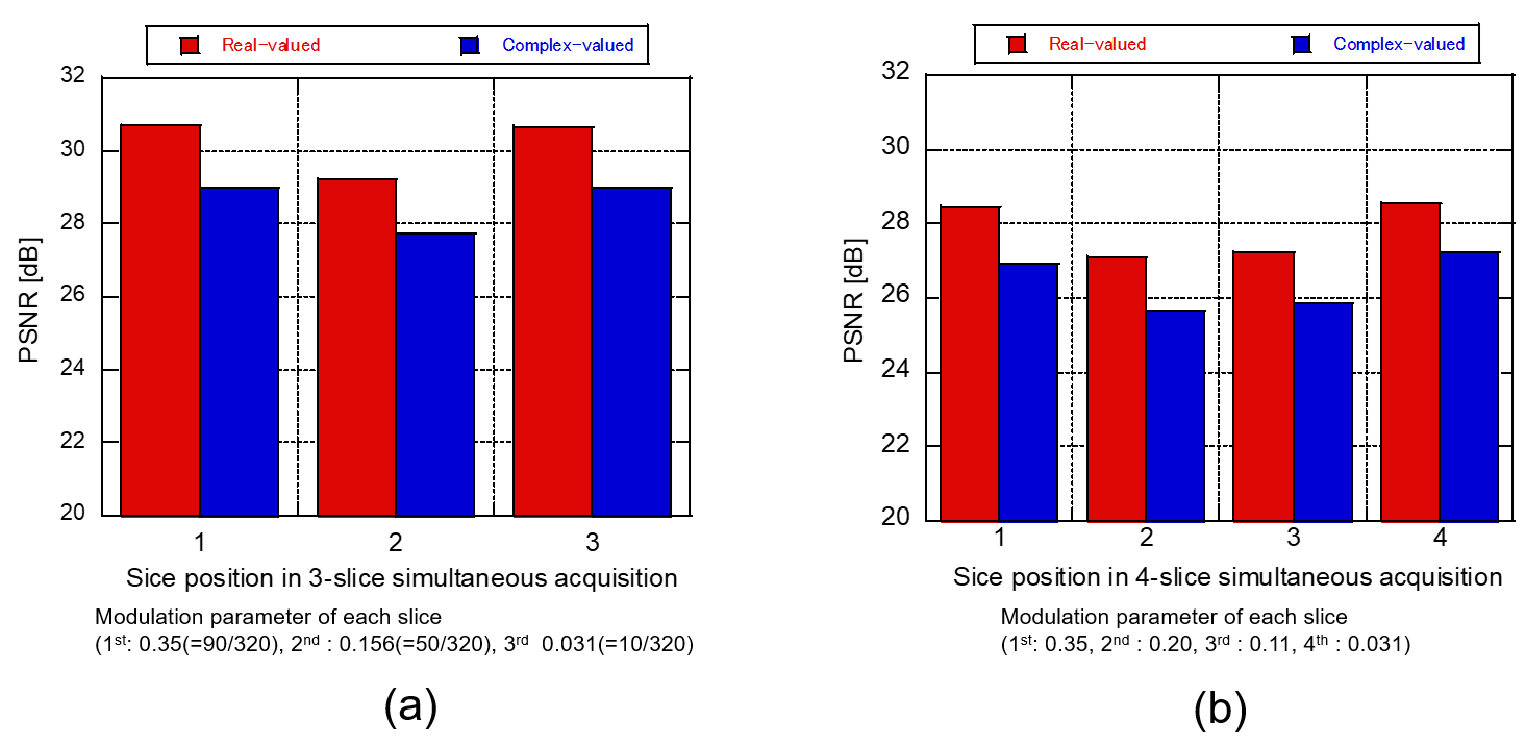}}
%
\vspace{-2mm}
\renewcommand{\baselinestretch}{0.3}
\caption{\small Relationship between PSNR and slice position. Relationship between slice position and PSNR for simultaneous (a) three-slice and (b) four-slice imaging.}
\label{fig:place}
\end{figure}
\subsection{Results of Fourier-holography-based SMS imaging}
Let the maximum frequency given by the sampling theorem be $k_{max}$. Then, the phase modulation coefficients can be expressed as $k_{mi} < \alpha_i k_{max} (\alpha_i <1), i=1,\ldots, n$. Figure 5(a) shows the relationship between the difference in phase modulation coefficients ($\alpha_3-\alpha_1)$ between the two end slices (slices 1 and 3) and the average PSNR for three-slice SMS imaging. $k_{m2}$ is set to the average of the values for the two ends ($ (k_{m1}+ k_{m3})/2 $). For training, 1400 T2-weighted images from the IXI dataset were used~\cite{4-1}.

Figure 5(b) shows the relationship between the number of slices excited simultaneously and the average PSNR when $ (\alpha_3-\alpha_1)$ is 0.79 ($=200/256$). For training, 400 T2-weighted images from the fastMRI dataset were used~\cite{4-2}.

%
\begin{figure}[t]
 \begin{center}
\centerline{\includegraphics[width=8.2cm]{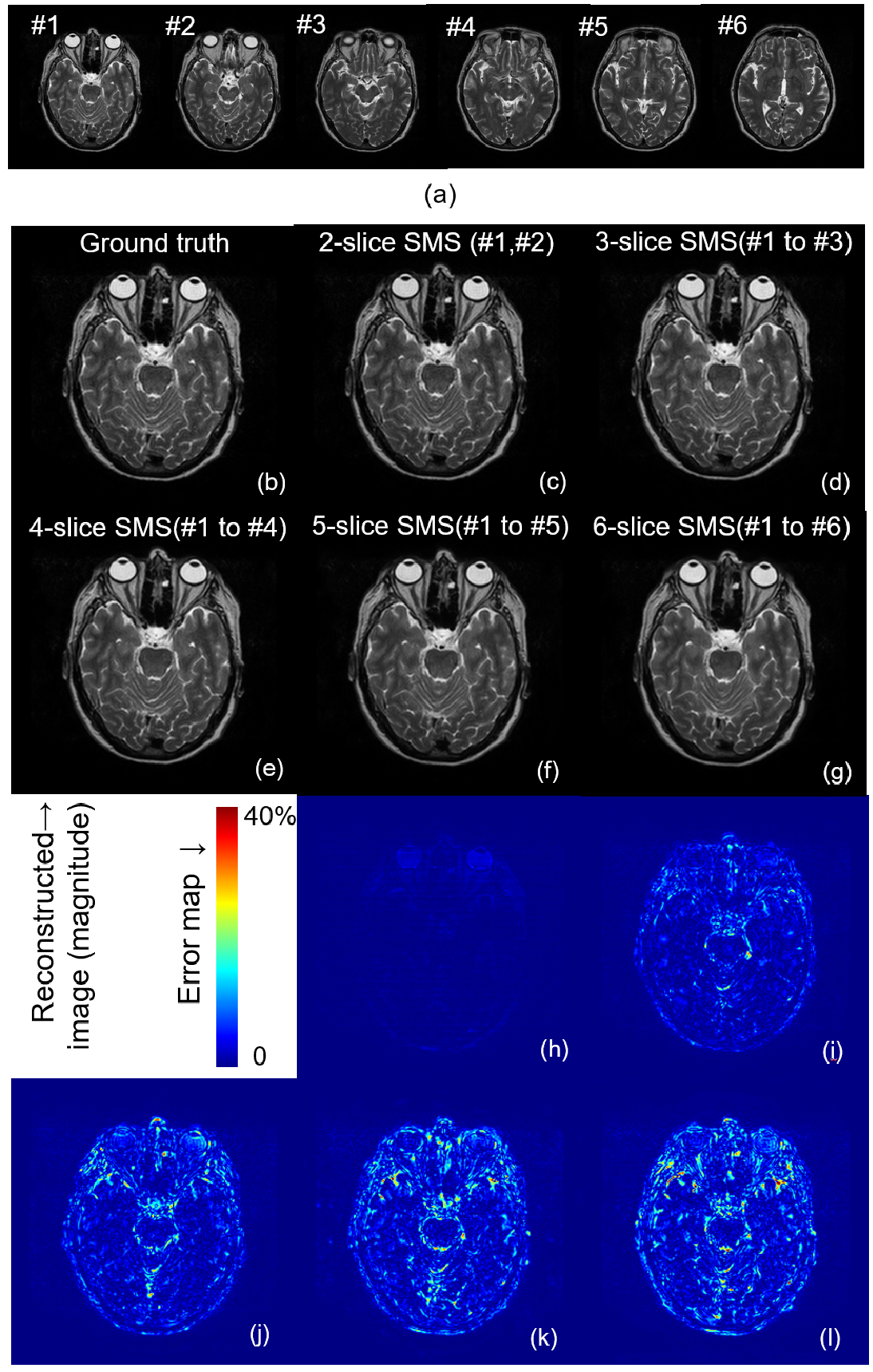}}
\renewcommand{\baselinestretch}{0.3}
  \caption{\small Reconstructed images of slice position 1 for various numbers of SMS slices.  (a)--(f) Six slice images used in simulation experiments, (g)--(k) reconstructed images at slice position 1, and (l)--(p) corresponding error images.}
  \label{}
 \end{center}
\end{figure}
Figure 6 shows the PSNR for each slice position for three- and four-slice SMS imaging. 
Figure 7 shows the reconstructed image of slice 1 for various numbers of simultaneously acquired slice images for the case of a real-valued image.
Figure 8 shows the simulation results for simultaneous three-slice excitation with a slice image that includes phase changes (which is assumed to be close to an actual situation). 

\subsection{Experiments using Fresnel-holography-based MR system}
Reconstructed images using Fresnel holography-based MR system are shown in Fig. 9.
Figure 9(a) shows the water phantom and (b) show the obtained MR signal (real part), (c) and (d) show temporally reconstructed images focused on the pin in foreground (right side) and focused on the far side (left side) using $\gamma b \tau$=1.44 and 1.56 ${\rm rad/cm^2}$ for $\gamma b \tau(1+z')$ term in Eq.(3), respectively. 
Pins in the focal plane are clearly visualized, but out-of-focus pins are blurred due to the diffraction effect of Fresnel transform.
Obtained images with proposed method corresponding to (c), (d) are shown in (e) and (f).
The results show that, only the pins of the focal plane are visualized.
\begin{figure}[t]
 \begin{center}
\centerline{\includegraphics[width=9cm]{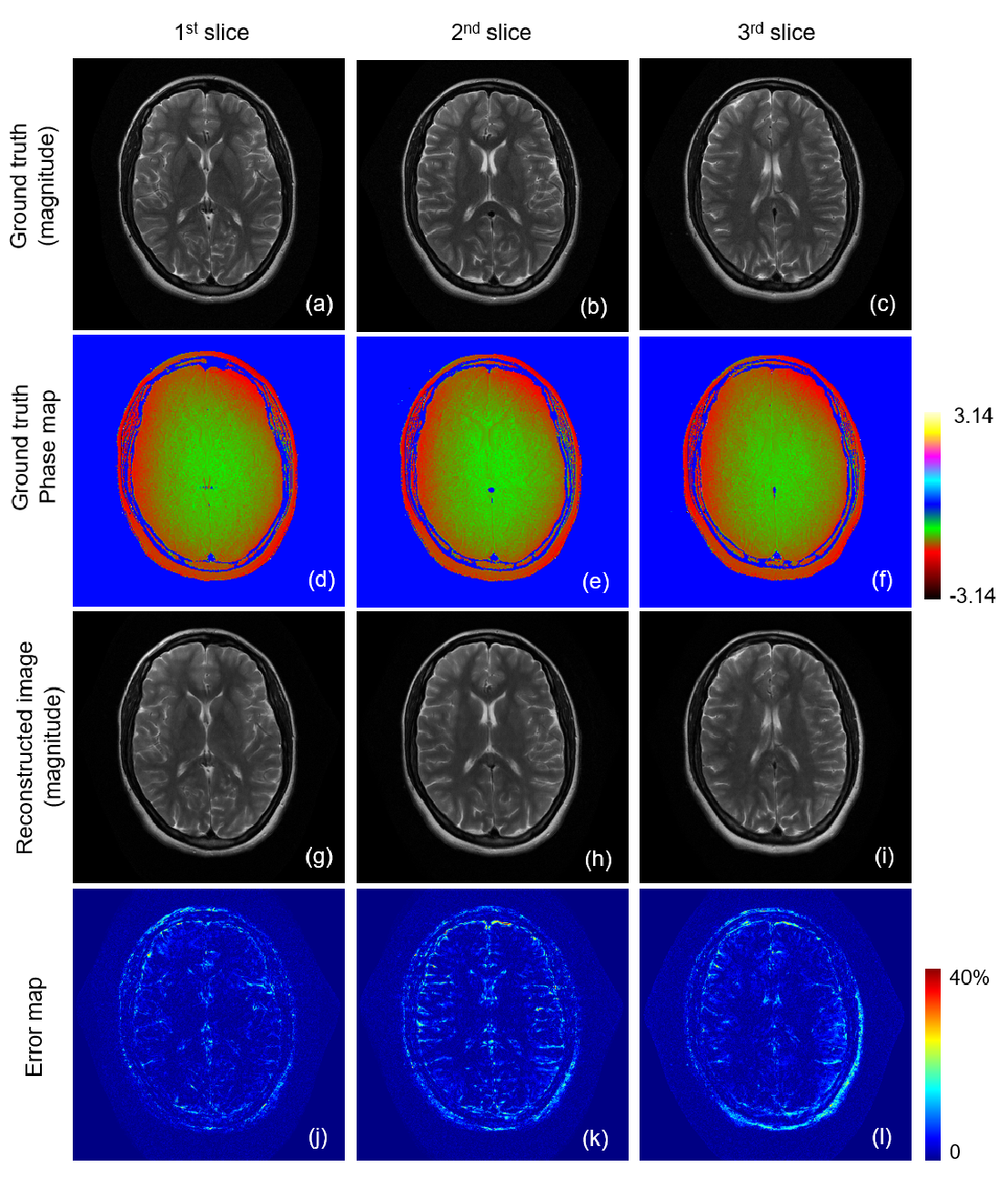}}
\renewcommand{\baselinestretch}{0.3}
  \caption{\small Simulation results of reconstructed images for simultaneous three-slice phase-varied image acquisition. (a)--(c) Fully scanned slice images, (d)--(f) phase map of each slice, (g)--(i) magnitude of reconstructed images, and (j)--(l) error images corresponding to (g)--(i), respectively.}
  \label{}
 \end{center}
\end{figure}
\section{Discussion}
\label{sec:discuss}
In the Fresnel hologram shown in Fig. 4, the out-of-focus image is diffracted. In the Fourier-holography-based shown in Fig. 3, the out-of-focus image is phase-modulated. Although the encoding formats of the out-of-focus image were different, both methods succeeded in removing the out-of-focus image.
Figure 5(a) suggests that the slice separation performance increases with increasing difference in the phase modulation coefficient.
This is because a larger phase modulation coefficient leads to a narrower stripe pattern for the out-of-focus image, as shown in Fig. 3, making it easier for the CNN to recognize the out-of-focus image.
\begin{figure}[t]
 \begin{center}
\centerline{\includegraphics[width=8cm]{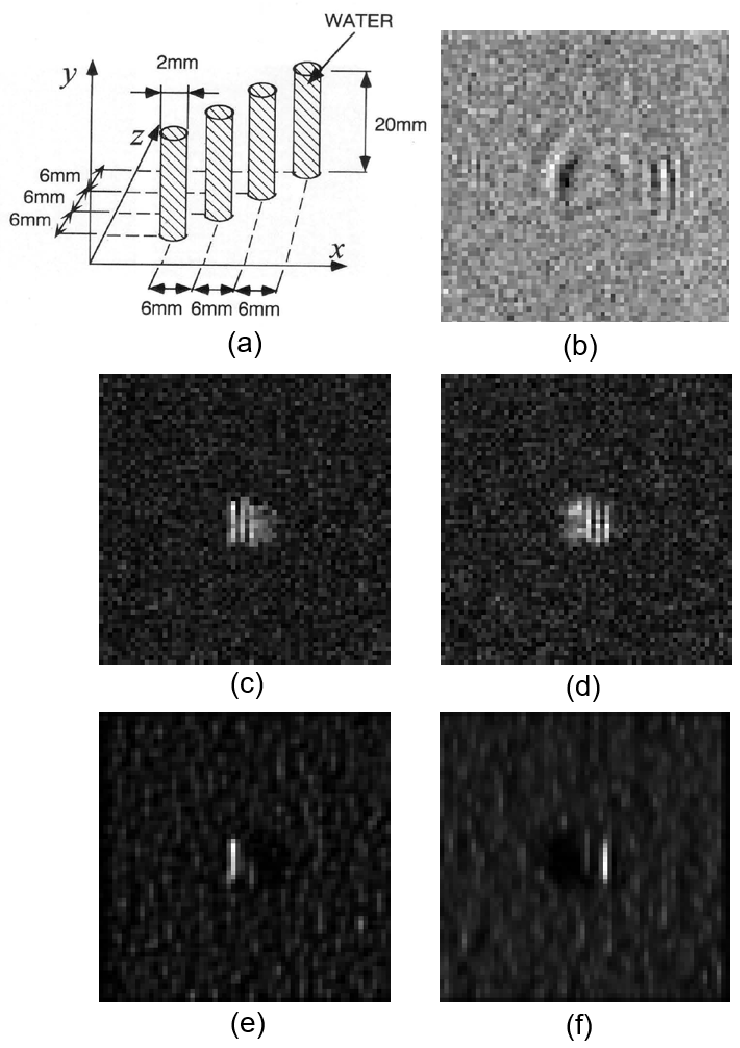}}
\renewcommand{\baselinestretch}{0.3}
  \caption{\small Reconstruction experiments using experimentally acquired Fresnel holography-based signal with 0.0183 T MRI.
(a) water phantom, (b) real part of acquired signal (64 $\times$ 64), (c) (d) temporally reconstructed images focused on the pins in foreground and on the far side, respectively, (e), (f) reconstructed images with proposed method.}
  \label{}
 \end{center}
\end{figure}

Figure 5(b) indicates that PSNR decreases with increasing number of simultaneously acquired slices.
Figure 7 shows the reconstructed image of slice 1 for two- to six-slice SMS imaging. It can be seen that the error in the contour of the image increases as the number of simultaneously acquired images increases. Figure 7 also shows that even with six-slice SMS imaging, most of the structure of the image is preserved.
This is due to the fact that as the number of slices simultaneously acquired increases, the focal-plane image component decreases and thus the error due to the removal of the out-of-focus image increases.
For images that contain phase changes, the PSNR is lower than that for real-valued images, but the results indicate that image separation is still possible.

Figure 6 shows that the PSNR is highest for the two end slices (i.e., slices closer to the center have lower PSNRs).
For three-slice SMS imaging, the difference in phase modulation coefficients between the center slice and the slices at the two ends is smaller than that between the slices at the two ends. As shown in Fig. 5(b), a greater difference in phase modulation coefficients leads to a better image separation performance. Therefore, the PSNR for the center slice is reduced.

Reconstruction experiments using phase-varied images showed a slight decrease in PSNR compared to that for real-valued images, but indicated that most of out-of-focus images were removed and image reconstruction is still possible. 
The proposed method uses spatial phase changes to separate the slice images of interest. Therefore, it is believed that phase changes that originate from MR equipment or living tissue will cause errors in image separation.

The MR signal used in the reconstruction experiments were those obtained by volumetric imaging~\cite{1-6}. Although imaging method was not a SMS, out-of-focus images were well removed and only the focal plane pins of interest could be visualized by approximating the volume of interests with a five-slice images.
Noises around the pins were also removed along with the out-of-focus pins, resulting in an almost noise-free blank area.
While the obtained  image quality is low, the effectiveness of the proposed method was demonstrated.

The proposed method is an ill-conditioned problem that cannot be solved using conventional methods. This research is an example of how deep learning can relax measurement conditions such as sampling theorems, and can be thought of as a new type of measurement that integrates measurement and deep learning.

In this study, the CNN was trained for each slice image. It is also possible to consider a method in which all slices are reconstructed with a single CNN.
Such a method is expected to provide higher image quality and is currently under consideration.
In this study, the basic performance of the proposed method was examined through simulation experiments. In the future, we plan to conduct SMS experiments using MRI.

%


\section{Conclusion}
\label{sec:conc}

We proposed an SMS imaging method that uses deep learning and the principle of holography.
Simulation experiments and initial MR imaging experiments show promising results for this method, which integrates measurement and reconstruction.
For the proposed Fourier-holography-based SMS imaging, image quality is improved when the coefficient of phase modulation is large. We clarified the relationship between the number of simultaneous images taken, spatial phase variation, and image quality. We plan to conduct experimental investigations in the future.



\bibliographystyle{IEEEbib}
\bibliography{strings,refs}

\end{document}